# An avalanche model for femtosecond optical response

J. B. Pendry[1]

[1] The Blackett Laboratory, Department of Physics, Imperial College London, London SW7 2BW, United Kingdom

**Abstract**
Experiments on Indium Tin Oxide (ITO) have revealed that a relatively broad 200THz pump pulse, roughly 30 cycles in width, can switch ITO from a transparent to a reflecting state within one or two optical cycles, a few Femtoseconds. So far the rapid switching has remained unexplained by theory. Here a very simple bare bones theory explains the key experimental features in terms of an avalanche model suggested by avalanche diodes.

**One-Sentence Summary**
An explanation is given for how reflectivity of a material can increase dramatically within one or two optical cycles.

**Main**

Indium tin oxide, an insulator[1], can be excited by a 200THz laser pump pulse, briefly to become a metal as electrons populate the conduction band[2]. The technique is widely used as a rapid switch for light by exploiting the transition from transparent dielectric to reflecting metal[3-5]. These and related experiments have driven the current interest in optical systems where the material parameters vary on a time scale comparable with one period of the radiation[6].

Recent experiments[7] reveal the rise time of reflectivity to be very short, almost two orders of magnitude shorter than previously thought, much shorter than the width of the pump pulse, and of the same order as one period of the pump: just a few femtoseconds. Experiments are now able to access the single cycle time scales that have been the subject of so much theoretical attention[6].

Current theories[8-11] have so far failed to explain this unsuspected rapid rise in reflectivity. Here we propose a mechanism derived by analogy with avalanche diodes[12]: an intense light field accelerates a background electron density and beyond a certain threshold these energetic electrons will initiate Auger processes exciting further electrons from the valence band. The events cascade into an avalanche with consequent exponential growth in electron density. Finally the electron plasma frequency exceeds that of any probe radiation resulting in reflection.

**The model**

Our model is designed to capture the key processes at work in the switching process. It eschews any extraneous complexity leaving visible the bare bones of this previously unexplained phenomenon. It is our hope that the model will inspire future experiments, enriching this rapidly growing field of optical experiments on time dependent materials.

Our starting point is data from a recent paper[7] which used a double-pulse pump to form two slits in time and thus, by analogy with Young's slits in space, creating fringes in frequency. Fringes extended far from the central frequency revealing a very short rise time of reflectivity. Both red and blue-shifted fringes were seen but showed massive asymmetry with red-shifted fringes far more intense than blue ones. Data were interpreted using an heuristic model which postulated a step function rise in reflectivity and reproduced the wide spread of frequency shifts in the red shift data but could not explain the asymmetry. This earlier model has no explanation for the origin of the step, an omission which we seek to remedy in this paper.

Our model must explain key aspects of experiment:

There is a catastrophic event causing extreme femtosecond time structure in reflectivity.

A threshold pump intensity is observed in experiment[7] of about 50 GW/cm$^2$.

Experiments show massive asymmetry between red- and blue- shifted interference fringes. If the reflectivity, $R(t)$, had no variation in phase with time, equal amplitudes for red and blue-shifted fringes would be inevitable therefore rapid change in phase must be a feature of any model.

The model is based on an avalanche diode which in back-bias mode passes no current until a threshold voltage is reached when residual conduction electrons are accelerated by the applied voltage to excite further electrons out of the valence band via an Auger process. The process cascades into an avalanche of electrons and the diode becomes highly conducting. In the case of an optical experiment we propose that electric fields in the pump also result in a sudden avalanche of electrons, provided that a minimum intensity is reached. It is well known that avalanches do not damage the diode which can recover after the voltage is removed, but on a longer time scale. As electron density increases so does the plasma frequency, $\omega_p$, and when $\omega_p > \omega_{\text{pump}}$ a sudden rise in reflectivity occurs along with dramatic changes in phase. Subsequent to the departure of the pump pulse, electrons relax into the valence band via electron-phonon interaction. The time scale for relaxation is known to be long and the model does not deal with this part of the problem.

**Detailed solution of the model**

As is the case in an avalanche diode we assume that there is a low density, $n_0$, of electrons present in the conduction band which can be accelerated. We shall show that this number is not critical to within many orders of magnitude, but we input typical values observed in dielectrics. An electric field $E$ delivers the following energy to an electron in time $t$,

$$U(t) = (Eet)^2/m \tag{1}$$

where $e$ is the electron charge and $m$ its effective mass. Our first assumption is that as soon as each electron has reached a threshold of energy, it excites a second electron across the band gap, $U_G$, via an Auger process. This energy is acquired after a time $t_G = \sqrt{U_G m}/Ee$. Recognising that fields in the pump reverse twice every period the process does not work unless electrons get enough energy to start an avalanche before the field reverses, that is to say in one half-cycle.

$$t_G = \sqrt{U_G m}/Ee < \pi/\omega_{pump} \qquad (2)$$

Assuming that as soon as enough energy is acquired an Auger process is triggered, the electron density obeys,

$$dn/dt = n/t_G = Eet/\sqrt{U_G m}\, n, \quad n = n_0 \exp\left(Eet/\sqrt{U_G m}\right) = n_0 \exp(\beta t) \qquad (3)$$

where $n_0$ is the background electron density. Assuming a band gap of $U_G = 3\text{eV}$ the model predicts a minimum power of,

$$U(t) = (Eet)^2/m = \left(Ee/(\omega_{pump}/\pi)\right)^2/m > U_G, \qquad (4)$$

implying cut-off at pump energies of roughly $10^{11}\,\text{W}/\text{cm}^2$, of the same order as the experimental observation of $5 \times 10^{10}\,\text{W}/\text{cm}^2$.

Next we turn to time evolution of the reflectivity. In a metal $\omega_p$ determines the permittivity which we crudely model as follows,

$$\varepsilon(\omega,t) = 1 - \omega_p^2/(\omega(\omega+i\gamma)) = 1 - n(t)e^2/(\omega(\omega+i\gamma)m\varepsilon_0) \qquad (5)$$

where $\gamma$ would normally be a positive quantity representing loss so that plasmon excitations decay in time. In contrast we are working with a system that is very strongly pumped. When $n$ is a function of time there is an additional contribution to $\gamma$,

$$\gamma = -\dot{n}/n = -\beta \qquad (6)$$

Evidence for the dominance of this term will come from the asymmetry of experimental spectra.

For simplicity we assume normal incidence of the probe giving an expression for reflectivity as a function of time,

$$R(t) = \frac{1-\sqrt{\varepsilon}}{1+\sqrt{\varepsilon}} = \frac{1-\sqrt{1-ae^{\beta t}}}{1+\sqrt{1-ae^{\beta t}}}, \quad a = \frac{n_0 e^2}{\omega(\omega+i\gamma)m\varepsilon_0} \qquad (7)$$

These are crude approximations but are sufficient to demonstrate the central features of our model. Refinement may be added in later works. Fortunately the model is insensitive to the parameters. Note $R(t)$ is almost real for $t < t_0$,

$$t_0 = -\beta^{-1} \ln \frac{n_0 e^2}{\omega^2 m \varepsilon_0} \qquad (8)$$

and complex with modulus almost unity for $t > t_0$.

**Predictions of the model – time dependence**

We use the following parameters to calculate reflectivity as a function of time, assuming a pump power of $10^{11}\,\text{W}(\text{cm})^{-2}$, just above threshold,

$$E \approx 10^9 \,\text{Vm}^{-1}, \qquad e = 1.6 \times 10^{-19} \,\text{C}, \quad U_G \approx 3\,\text{eV}$$
$$m = 9.11 \times 10^{-31} \,\text{kg}, \quad f_{\text{pump}} = 200\,\text{THz}, \quad n_0 = 10^{24} \,\text{m}^{-3} \tag{9}$$

Fig. 1 confirms that our model predicts a 1-2 cycle rise time for reflectivity, which will give extended structure in fringe amplitudes after Fourier transformation. Calculating for two different values of the gain parameter shows relative insensitivity. The larger figure would imply that the plasmon amplitude was more than doubling over ten cycles. Later figures assume the smaller value.

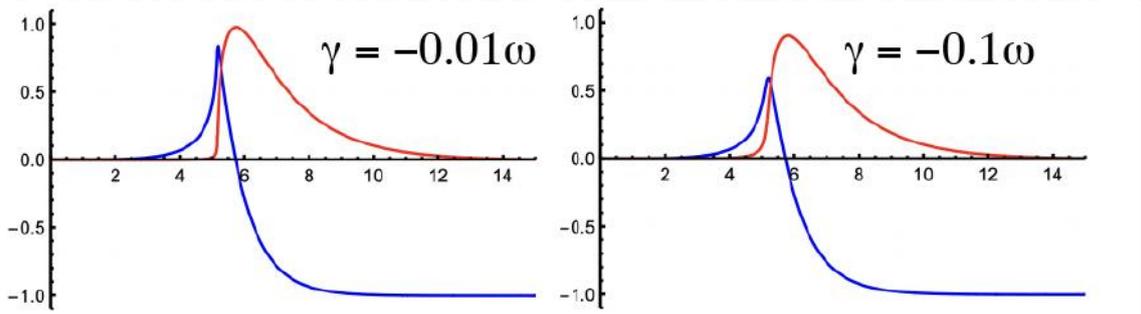

**Fig. 1** Real (blue) and imaginary (red) parts of $R(t)$ plotted as a function of time in units of the pump period for a background conduction electron density of $n_0 = 10^{24} \,\text{m}^{-3}$. Left: calculated for $\gamma = -0.01\omega$; right: for $\gamma = -0.1\omega$.

Our assumption of a background electron density of $n_0 = 10^{24} \,\text{m}^{-3}$ is not a critical one as can be seen from Fig. 2: reducing $n_0$ by a factor of $10^{-2}$ simply delays the rise in reflectivity; the pump has to work for longer to attain the critical density. However the form of the rise in reflectivity does not change as can be seen in Fig. 2b where the reduced density curve has been translated backwards in time to show its perfect overlap with the higher density curve.

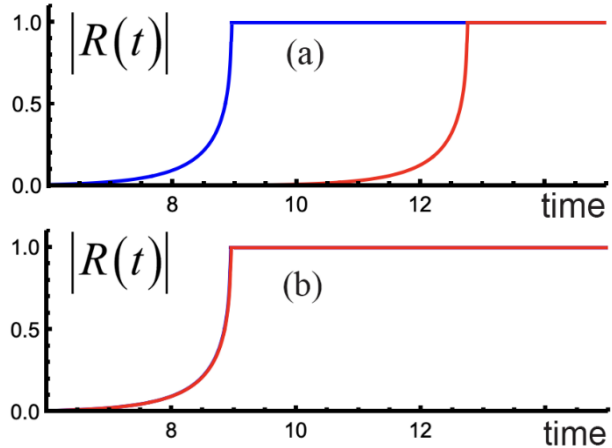

Fig. 2 (a) blue: as for Fig. 1 but absolute values of reflectivity plotted, $n_0 = 10^{24} \,\text{m}^{-3}$; red: now with $n_0 = 10^{22} \,\text{m}^{-3}$. (b) the red curve is shifted backwards in time to show that it has the same profile, overlapping exactly with the blue curve.

Next we investigate dependence on the pump power. In Fig. 3 we plot $|R(t)|$ as a function of time in units of the pump period for a pump power of $10^{13} \text{W}(\text{cm})^{-2}$ for the leftmost curve. For subsequent curves power is reduced by factors of 0.16, 0.04, 0.01 the latter corresponding to $10^{11} \text{W}(\text{cm})^{-2}$. Lower power implies later onset of the avalanche. A background conduction electron density of $n_0 = 10^{20} \text{m}^{-3}$ is assumed. As is evident from the figure, higher power compensates for much lower background density, once more demonstrating insensitivity to choice of parameters. However note that higher power means an increased pre-factor in the exponent with time, shortening the rise time to less than one pump period.

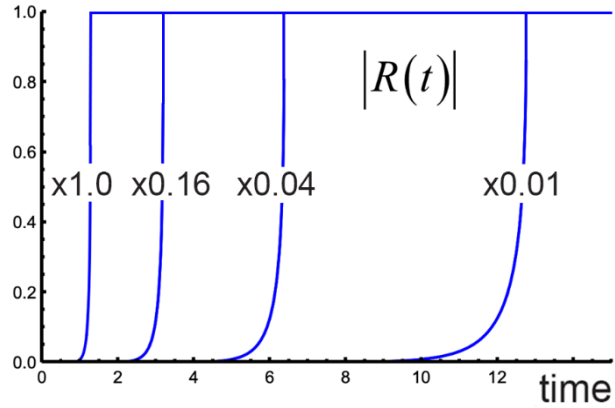

Fig. 3 $|R(t)|$ as a function of time in units of the pump period for various pump powers starting at $10^{13} \text{W}(\text{cm})^{-2}$.

**Predictions of the model – frequency dependence**

Experiments showing extended fringe visibility as a function of frequency shift implied rapid rise of reflectivity with time. To calculate visibility we Fourier transform $R(t)$ to obtain $R(\delta)$.

Fourier transformation of $R(t)$,

$$R(\delta) = \int_0^\infty R(t) e^{i\delta t} dt = \int_0^\infty \frac{1 - \sqrt{1 - ae^{\beta t}}}{1 + \sqrt{1 - ae^{\beta t}}} e^{i\delta t} dt$$

$$= \int_{-\infty}^\infty \frac{1 - \sqrt{1 - ae^{\beta t}}}{1 + \sqrt{1 - ae^{\beta t}}} e^{i\delta t} dt - \int_{-\infty}^0 \frac{1 - \sqrt{1 - ae^{\beta t}}}{1 + \sqrt{1 - ae^{\beta t}}} e^{i\delta t} dt$$

(10)

is a subtle process requiring consideration of the analytic structure of $R(t)$ which contains a branch point located in the upper half plane if $\gamma > 0$, or the lower half plane if $\gamma < 0$. As discussed above the latter case holds in the presence of pumping. Fig. 4 shows the strategy for manipulating the contour of integration.

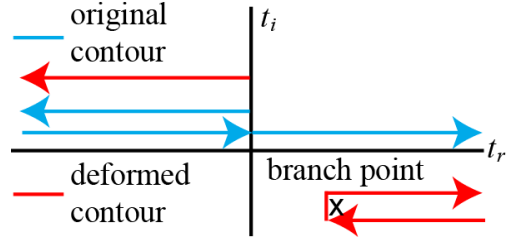

Fig. 4 Contours of integration in the complex time plane showing the original contours in blue and the distorted contours in red for negative values for $\delta < 0$, in which case the contours can be closed in the lower half plane. When $\delta > 0$ the contour must be closed in the upper half plane and the branch cut is omitted.

After readjusting the contour,

$$R_-(\delta) = \int_{t_0}^{\infty} \frac{-4\sqrt{1-ae^{\beta t}}}{ae^{\beta t}} e^{i\delta t} dt - \int_{-\infty}^{0} \frac{1-\sqrt{1-ae^{\beta t}}}{1+\sqrt{1-ae^{\beta t}}} e^{i\delta t} dt, \quad \delta < 0,$$

$$R_+(\delta) = -\int_{-\infty}^{0} \frac{1-\sqrt{1-ae^{\beta t}}}{1+\sqrt{1-ae^{\beta t}}} e^{i\delta t} dt, \quad \delta > 0$$

(11)

Using these formulae we calculate fringe visibility shown in Fig. 5 plotted against frequency shift. Experimental data in an earlier paper[7] measured fringes extending to shifts of the order of 20% of the probe frequency.

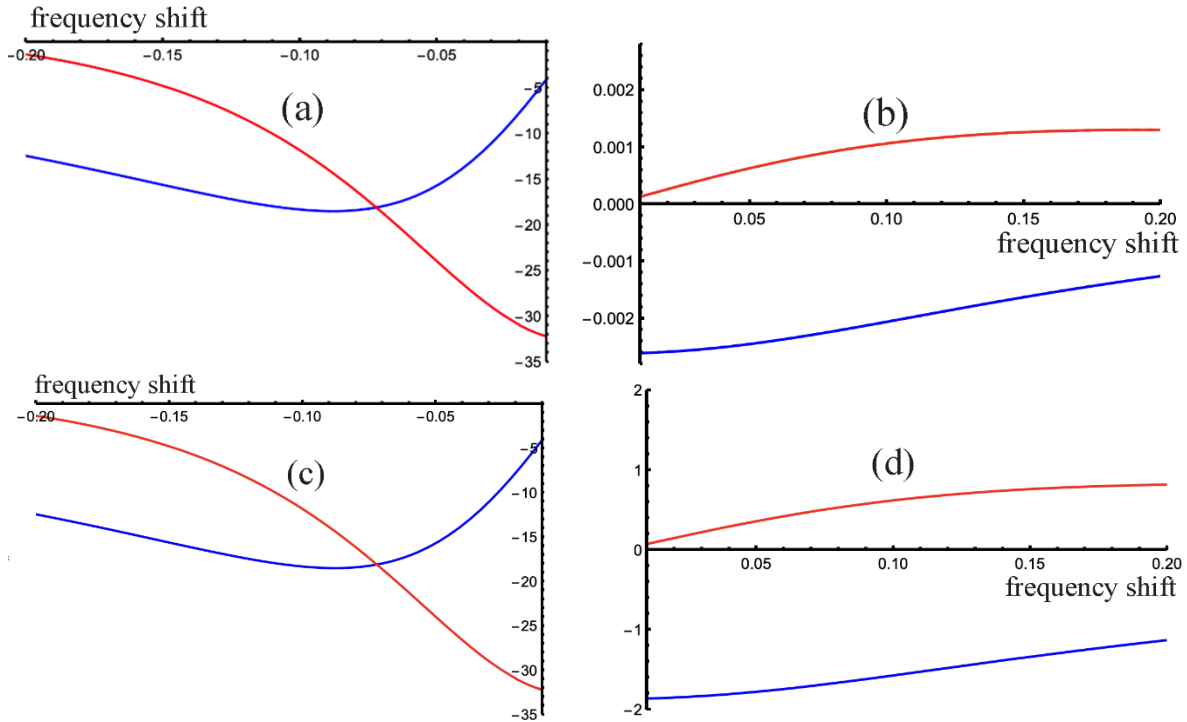

Fig. 5 (a) Real (blue) and imaginary (red) parts of $R_-(\delta)$ and (b) $R_+(\delta)$ in units of $\omega_{\text{pump}}^{-1}$, with $\delta$ shown as a fraction of the probe frequency. The assumed background conduction electron density is , $t_0 = 2.6 \times 10^{-14}$, $n_0 = 10^{24} \text{m}^{-3}$. (c) and (d) as for (a) and (b) but with $t_0 = 0$, $n_0 = 4.8 \times 10^{26} \text{m}^{-3}$.

Fig. 5 shows that our model can explain the high visibility of fringes extending to a substantial fraction of the probe frequency. It also accounts for the dramatic asymmetry between positive and negative frequency shifts. In fact because the contribution from $t < 0$ is very small we can approximate,

$$R_+(\delta) \approx \frac{-n_0 e^2}{4\omega_{\text{pump}}^2 m\varepsilon_0 (\beta + i\delta)} \tag{12}$$

and the ratio between $R_+$ and $R_-$ can be used to estimate $n_0$. In some experiments the background conduction electron density is tuned to give an 'epsilon near zero' at the pump frequency in order to enhance the pump power and as a result the system is already at the onset of negative permittivity, $t_0 = 0$, and the reflecting state occurs substantially within the first few cycles of the probe. Figs. (5c,d) show that, as predicted by eq. (12), the blue shifted fringes, $R_+(\delta)$, are now much stronger than at lower values of $n_0$ but still substantially weaker than those red shifted. Again this is in qualitative agreement with experiment.

Although Fig. 2 shows that $|R(t)|$ approximates to the step function postulated in the earlier model[4], Fig 1 adds more to the story. When considering the phase of as well as the amplitude of $R(t)$ Fig. 1 reveals that $\text{Im} R(t)$ is more like a delta function than a step and in contrast to a step would predict a fringe amplitude constant in frequency shift. There is evidence of this in the experiments: whereas the step model predicts a constant amplitude for $|\delta \times R_-(\delta)|^2$ experiments observe a slight initial rise in this quantity which was puzzling at the time but is explained by the present model.

**Summary of the model calculations**

We set our model three challenges and all of these have been met:

Our model reproduces extreme femtosecond time structure as evidenced by Fig. 1. Fringe visibility is shown in Fig 5 to extend at least as far as observed in experiment. The 'catastrophe' referred to is revealed as an avalanche driving up electron density at an exponentially increasing rate hence leading to a very rapid onset of reflectivity.

Our model exhibits a threshold pump intensity calculated to be of the same order as in experiment.

The dramatic asymmetry of fringes between positive and negative frequency shifts is explained by analytic structure in the model. Asymmetry can be used to measure the background electron density which triggers the avalanche.

**Thoughts on future experiments**

Existing experimental data are in agreement with our model, but in the light of our further predictions it would be useful to see more data. For example theory predicts that lower pump power and/or lower $n_0$ would shift $t_0$ to later in the pulse. This could be measured in a double slit experiment by introducing different intensities for the two pump pulses and thus shifting the periodicity of the interference fringes which is easily and accurately measured. Does our theory make a correct prediction?

In fact the red shifted fringes depend strongly on pump intensity as shown below. Increasing the power results in sharper structure in time which in turn spreads the Fourier components over a wider range of frequencies but with lower amplitude at a given frequency.

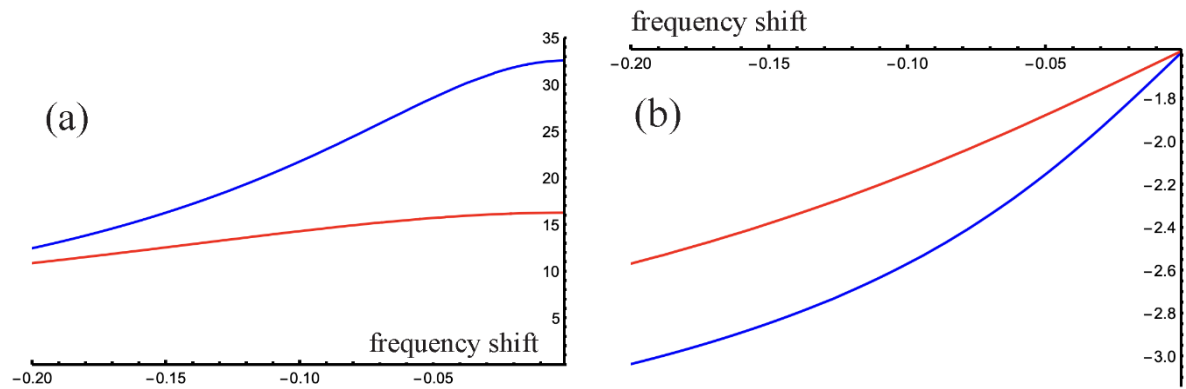

Fig. 6 (a) the amplitude of $R_-(\delta)$, $\delta$ in units of $\omega_{pump}^{-1}$, and (b) the phase in radians. Red curves: pump power of $2\times 10^{13} \text{W}(\text{cm})^{-2}$; blue curves $10^{13} \text{W}(\text{cm})^{-2}$ with $\delta$ shown as a fraction of the probe frequency. The assumed background conduction electron density is, $t_0 = 2.6\times 10^{-14}$, $n_0 = 10^{24} \text{m}^{-3}$.

Imbalance between red and blue shifted fringes depends strongly on both $n_0$ and $\beta$, the latter determined by the pump power. Measuring the imbalance between, say the 5$^\text{th}$ red and 5$^\text{th}$ blue shifted fringes as a function of these parameters would test the theory.

The earlier empirical theory[7] postulated a step in reflectivity but we now have a more nuanced model. Taking measurements of fringes to higher red shifts would reveal departures from the step model. Also fig. 5 predicts significant changes in phase with frequency shift. Could the phase of the fringes be measured by interference with a reference signal?

Our model for rapid switching of reflectivity is inspired by an avalanche diode which raises the question of whether the limiting action of an actual diode could give rise to the same switching thus opening the way to cheap and efficient pure electrical control of the reflectivity. Optical switching has a threshold of around $10^{11} \text{W}(\text{cm})^{-2}$, or field strengths of around $10^9 \text{Vm}^{-1} = 10^3 \text{V}\mu^{-1}$ which is not far from the operating range of avalanche diodes. Driven at a few hundred GHz a diode would produce a saw-tooth like profile in reflectivity, driven by the rapid rise of the avalanche followed by slower relaxation as electrons and holes recombine.

**Funding**

I acknowledge support from EPSRC reference EP/Y015673/1

**Acknowledgements**

I thank our experimental team led by Riccardo Sapienza for extensive discussion of their experiments, and Simon Horsley of some theoretical points.